% Title:   Janossy Densities, multimatrix Spacing distributions
%          and Fredholm determinants
% Authors: J. Harnad 
% Type:    Preprint CRM-2978 (2004)
% Date:    March 1, 2004
% ===========================================================
% Compiler: Plain TeX
% ===========================================================
%%%%%%%%%%%%%%%%%%%%%%%%%%%%%%%%%%%%%%%%%%%%%%%%%%%%%%%%%%%%%%%%%%%%%%%%%%%%
%
% (Macros in Plain TeX by Aurel Wisse, Version 1.4)
%
%%%%%%%%%%%%%%%%%%%%%%%%%%%%%% Useful Definitions %%%%%%%%%%%%%%%%%%%%%%%%%
%
\def\@{{\char'100}}

\long\def\abstract#1{\bigskip{\advance\leftskip by 2true cm
\advance\rightskip by 2true cm\eightpoint\centerline{\bf
Abstract}\everymath{\scriptstyle}\vskip10pt\vbox{#1}}\bigskip}
\long\def\resume#1{{\advance\leftskip by 2true cm
\advance\rightskip by 2true cm\eightpoint\centerline{\bf
R\'esum\'e}\everymath{\scriptstyle}\vskip10pt \vbox{#1}}}

\def\references{\bigbreak\centerline{\sc
References}\medskip\nobreak\bgroup
\def\ref##1&{\leavevmode\hangindent45pt
\hbox to 42pt{\hss\bf[##1]\ }\ignorespaces}
\parindent=0pt
\everypar={\ref}\par}
\def\endreferences{\egroup}
\long\def\authoraddr#1{\medskip{\baselineskip9pt\let\\=\cr
\halign{\line{\hfil{\Addressfont##}\hfil}\crcr#1\crcr}}}

%
% Runningheads
%
\newif\ifrunningheads
\runningheadstrue \immediate\write16{- Page headers}
\headline={\ifrunningheads\ifnum\pageno=1\hfil\else\ifodd
\pageno\rightheadline
\else\leftheadline\fi\fi\else\hfil\fi}
\def\rightheadline{\sc\hfil\RightHeadText\hfil}
\def\leftheadline{\sc\hfil\LeftHeadText\hfil}

\hyphenation{Harnad Neumann}
%
%%%%%%%%%%%%%%%%%%%%%%%%%%%%%%%% Fonts %%%%%%%%%%%%%%%%%%%%%%%%%%%%%%%%%%%%%
%
\immediate\write16{- Fonts "Small Caps" and "EulerFraktur"}
%
%
%%  The following two lines introduce the Small Caps font at 10pt,
%%  if not available comment out and replace by the
%%  following line:
% \def\sc{\rm}
%

\let\sc=\tensmc
%
%%  The following four lines introduce the Euler Fraktur font,
%%  if not available comment out and replace by
%%  the line:
% \def\gr{}
%
\font\teneuf=eufm10  \font\seveneuf=eufm7 \font\fiveeuf=eufm5
\newfam\euffam 

\textfont\euffam=\teneuf \scriptfont\euffam=\seveneuf
\scriptscriptfont\euffam=\fiveeuf
%
%%%%%%%%%%%%%%%%%%%%%%%%%%%% Definitions %%%%%%%%%%%%%%%%%%%%%%%%%%%%%%%%

\def \wt {\widetilde}

\def \ra {\rightarrow}

\def \lra {\longrightarrow}

\def \d {\delta}

\def \ss {\subset}

\def \det{{\rm det}}

\def\nchi{\hbox{\raise 2.5pt\hbox{$\chi$}}}
%
%Euler Fraktur letters
%

\def\nchi{\hbox{\raise 2.5pt\hbox{$\chi$}}}
%
% Calligraphic roman letters for Plain Tex
% for Plain Tex
%

%
%Bold roman letters
%

%
% Title, Author, Address and Abstract fonts
%
\def\authorfont{\sc}
\font\eightrm=cmr8 \font\eightbf=cmbx8 \font\eightit=cmti8
\font\eightsl=cmsl8 
\def\eightpoint{\let\rm=\eightrm \let\bf=\eightbf \let\it=\eightit
\let\sl=\eightsl \baselineskip = 9.5pt minus .75pt  \rm}

\font\titlefont=cmbx10 scaled\magstep2 \font\sectionfont=cmbx10
 \font\Addressfont=cmsl8
%
% Theorems
%
\def\Proclaim#1:#2\par{\smallbreak\noindent{\sc #1:\ }
{\sl #2}\par\smallbreak}
\def\Demo#1:#2\par{\smallbreak\noindent{\sl #1:\ }
{\rm #2}\par\smallbreak}
%
% Section headings
%
\immediate\write16{- Section headings}
\newcount\secount
\secount=0
\newcount\eqcount
\outer\def\section#1.#2\par{\global\eqcount=0\bigbreak \ifcat#10
 \secount=#1\noindent{\sectionfont#1. #2}
\else
 \advance\secount by 1\noindent{\sectionfont\number\secount. #2}
\fi\par\nobreak\medskip}
%
% Definition of automatic numbering
% and corresponding commands
%
\immediate\write16{- Automatic numbering} \catcode`\@=11
\def\adv@nce{\global\advance\eqcount by 1}
\def\unadv@nce{\global\advance\eqcount by -1}
\def\nextnumber{\adv@nce}
%
% Automatic numbering
%
\newif\iflines
\newif\ifm@resection
\def\onesec{\m@resectionfalse}
\def\moresec{\m@resectiontrue}
\moresec
\def\eq{\global\linesfalse\eq@}
\def\eqn{\global\linestrue&\eq@}
\def\nosubind@x{\global\subind@xfalse}
\def\newsubind@x{\ifsubind@x\unadv@nce\else\global\subind@xtrue\fi}
\newif\ifsubind@x
\def\eq@#1.#2.{\adv@nce
 \if\relax#2\relax
  \edef\loc@lnumber{\ifm@resection\number\secount.\fi
  \number\eqcount}
  \nosubind@x
 \else
  \newsubind@x
  \edef\loc@lnumber{\ifm@resection\number\secount.\fi
  \number\eqcount#2}
 \fi
 \if\relax#1\relax
 \else
  \expandafter\xdef\csname #1@\endcsname{{\rm(\loc@lnumber)}}
  \expandafter
  \gdef\csname #1\endcsname##1{\csname #1@\endcsname
  \ifcat##1a\relax\space
  \else
   \ifcat\noexpand##1\noexpand\relax\space
   \else
    \ifx##1$\space
    \else
     \if##1(\space
     \fi
    \fi
   \fi
  \fi##1}\relax
 \fi
 \eq@@{\loc@lnumber}}
\def\eq@@#1{\iflines \else \eqno\fi{\rm(#1)}}
\def\m@th{\mathsurround=0pt}
%
% Command \display
%
\def\display#1{\null\,\vcenter{\openup1\jot
\m@th \ialign{\strut\hfil$\displaystyle{##}$\hfil\crcr#1\crcr}}
\,}
\newif\ifdt@p
\def\@lign{\tabskip=0pt\everycr={}}
\def\displ@y{\global\dt@ptrue \openup1 \jot \m@th
 \everycr{\noalign{\ifdt@p \global\dt@pfalse
  \vskip-\lineskiplimit \vskip\normallineskiplimit
  \else \penalty\interdisplaylinepenalty \fi}}}
%
% Command \displayno
%
\def\displayno#1{\displ@y \tabskip=\centering
 \halign to\displaywidth{\hfil$
\@lign\displaystyle{##}$\hfil\tabskip=\centering&
\hfil{$\@lign##$}\tabskip=0pt\crcr#1\crcr}}
%
% References
%
\def\cite#1{{[#1]}}
\catcode`\@=\active
%
%%%%%%%%%%%%%%%%%%%%%%%%%%%%%%%% Formatting %%%%%%%%%%%%%%%%%%%%%%%%%%%%
%
\magnification=\magstep1 \hsize= 6.75 true in \vsize= 8.75 true in
%
%%%%%%%%%%%%%%%%%%%%%%%%%%%%%%%%%%% Headers %%%%%%%%%%%%%%%%%%%%%%%%%%%%%%
%
\def\RightHeadText{Janossy densities and Fredholm resolvents}
\def\LeftHeadText{J. Harnad }
%
%%%%%%%%%%%%%%%%%%%%%%%%% Document %%%%%%%%%%%%%%%%%%%%%%%%%%%%%%%%%%%%%%
%
\rightline{CRM-2978 (2004) \break} 
%\rightline{solv-int/0401xx\break} 
\bigskip 
\centerline{\titlefont Janossy densities, multimatrix spacing distributions}
\centerline{\titlefont{and Fredholm resolvents}}
%\footnote{${}^{\dagger}$}{\eightpoint
%Research supported in part by the Natural Sciences and
%Engineering Research Council of Canada and the Fonds FCAR du Qu\'ebec.}}
\bigskip
\centerline{\authorfont J.~Harnad}
\authoraddr
{Department of Mathematics and Statistics, Concordia University\\
7141 Sherbrooke W., Montr\'eal, Qu\'ebec, Canada H4B 1R6, {\rm
\eightpoint
and} \\
Centre de recherches math\'ematiques, Universit\'e de Montr\'eal\\
C.~P.~6128, succ. centre ville, Montr\'eal, Qu\'ebec, Canada H3C 3J7\\
{\rm \eightpoint e-mail}: harnad\@crm.umontreal.ca}
\bigskip

\bigskip
\abstract{A simple proof is given for a generalized form of a theorem of
Soshnikov. The latter states that the Janossy densities for multilevel
determinantal ensembles supported on measurable subspaces of a set of measure
spaces are constructed by dualization of bases on dual pairs of
$N$-dimensional function spaces with respect to a pairing given by
integration on the complements of the given measurable subspaces. The
generalization extends this to dualization with respect to measures modified
by arbitrary sets of weight functions.}
\bigskip
\baselineskip 14 pt

%%%%%%%%%%%%%%%%%%%%%%%%%%%% Section 1. Introduction %%%%%%%%%%%%%%%%%%%
\section 1. Multilevel determinantal ensembles.
\smallskip \nobreak

Let $\{(\Gamma_j, d\mu_j)\}_{j=1 \dots m}$ be a set of measure spaces and
$\{\tilde{H_j}:= L^2(\Gamma_j, d\mu_j)\}_{j=1 \dots m}$ the Hilbert spaces
of square integrable functions on them. Let $\{f_a\}_{a=1 \dots N}$ and
$\{h_a\}_{a=1 \dots N}$ be bases for a pair of $N$-dimensional subspaces
$H_1 \ss \tilde{H}_1$, $H^m \ss \tilde{H}_m$,  respectively. Suppose we 
are given $m -1$ functions  $\{g_{j+1, j}\}_{j=1\dots m-1}$  on
the product spaces $\Gamma_{j+1}\times \Gamma_{j}$, such that the 
corresponding integral operators
$$
\eqalign{g_{j+1, j}: \hat{H}_j &\lra  \hat{H}_{j+1} \cr
g_{j+1, j}(f)(x_{j+1}) :=&  \int_{\Gamma_j} g_{j+1,j}(x_{j+1},
x_j)f(x_j)d\mu_j(x_j),}
\eq ..
$$
together with their transposes
$$
\eqalign{g^*_{j+1, j}: \hat{H}_{j+1} &\lra  \hat{H}_j \cr
g^*_{j+1, j}(f)(x_{j}) :=&  \int_{\Gamma_{j+1}} g_{j+1,j}(x_{j+1},
x_j)f(x_{j+1})d\mu_j(x_{j+1}),}
\eq..
$$
are well defined injective maps on a sequence of dense subspaces
$\hat{H}_1\ss \wt{H}_1$ , $\hat{H}_2 = g_{21}(\hat{H}_1)\ss \wt{H}_2$,
$\dots,$
$\hat{H}_m =  g_{m,m-1}(\hat{H}_{m-1})\ss\wt{H}_m$, as are their composites:
$$
g_{k j} := g_{k,k-1} \circ \dots \circ g_{j+1,j}.
\eq gijint..  
$$
We assume  that $H_1\ss \hat{H}_1$, $H^m\ss \wt{H}_m$,  and denote the
respective images as $H_2:= g_{21}(H_1), \dots , $ $H_m:= g_{m, m-1}(H_{m-1})$
and  $H^{m-1}:= g^*_{m, m-1}(H^m), \dots ,$ $H^1:= g^*_{2, 1}(H^2)$.
Assuming furthermore that the $N \times N$ matrix:
$$
A_{ab} := \int_{\Gamma_1} f_a(x_1) g^*_{m1}(h_b)(x_1) d\mu_1(x_1) = 
\int_{\Gamma_m} g_{m1}(f_a)(x_m) (h_b)(x_m) d\mu_m(x_m) 
\eq Apairing..
$$
is nonsingular, it follows that the pairs of spaces $\{H_j, H^j\}$ may be 
viewed as mutually dual, and identified through the $d\mu_j$ integration
pairings. Making a $PLU$ decomposition of $A$ 
$$
A :=  P L U,  \eq..
$$
where $L$ and $U$ are, respectively, lower and upper triangular matrices,
normalized, e.g., with equal diagonal entries (unique up to $2^N$ sign
ambiguities on the diagonal), and $P$ a permutation matrix, we may form 
bases:
$$
\psi_a^{(1)} := \sum_{b=1}^N (PL)^{-1} _{ab} f_b \qquad \phi_a^{(m)} :=
\sum_{b=1}^N U^{-1} _{ba} h_b 
\eq..
$$
for the spaces $H_1$ and $H^m$, respectively, as well as for the
sequence of dual spaces $\{H_j, H^j\}$ through composition with $g_{j1}$ and
$g^*_{mj}$:
$$
\{\psi_a^{(j)}:=g_{j1} (\psi_a^{(1)})\}_{a=1\dots N}, \qquad
\{\phi_a^{(j)}:=g^*_{mj} (\phi_a^{(m)})\}_{a=1\dots N}.
\eq..
$$
These are, by construction, mutually dual,
$$
\int_{\Gamma_j} \psi_a^{(j)}(x_j) \phi_b^{(j)}(x_j) d\mu_j(x_j) = \delta_{ab}.
\eq dualpsiphi..
$$

An example of the above construction consists of choosing the sets
$\Gamma_j$ to be intervals on the real line, with Lebesgue measure, and the
functions $f_a, h_a$ and $\{g_{j,j-1}\}$ of the form:
$$
\eqalign{
f_a(x_1) :=x_1^{a-1} e^{-{1\over 2} V_1(x_1)}, &\quad  h_a(x_m): = x_m^{a-1}
e^{-{1\over 2} V_m(x_m)},\cr
g_{j, j-1}(x_j, x_{j-1})&:=e^{ x_{j-1} x_{j} - {1\over 2}(V_{j-1}(x_{j-1})
+V_{j}(x_j))},} \eq..
$$
where the $V_j(x_j)$'s are suitably defined functions for which the integrals
involved are convergent (e.g., polynomials of even degree with real, positive
leading coefficients).  This case arises in the study of random multimatrix
chains models [EM, BEH1, BEH2]. The reduced probability density for 
eigenvalues
$\{x^{(j)}_a\}_{j=1
\dots m, a=1
\dots N}$ is given by the determinantal formula
$$
P_N^m (x^{(j)}_a) = {1\over Z_{N,m}} \det(\psi^{(1)}_a(x^{(1)}_b))
\det(\phi^{(m)}_a(x^{(m)}_b))\prod_{j=1}^{m-1} \det (g_{j+1, j}(x^{(j+1)}_a,
x^{(j)}_b))
\eq prob..
$$
where the partition function $Z_{N,m}$ is defined so as to normalize this 
to a probablility measure. 

More generally, let us suppose that the functions $f_a, h_a, g_{j, j-1}$ are
chosen so that the expression \prob defines a probablity measure on the 
ensemble $\prod_{j=1}^m (\Gamma_j)^ N$. Such ensembles are sometimes referred to
as ``determinantal ensembles'' [BS, S]; other examples include, e.g., 
``polynuclear growth'' models [PS, J]. 

Define the following set of functions on the product spaces 
$\Gamma_i \times \Gamma_j$:
$$
K_{ij}(x_i, x_j) := \sum_{a=1}^N \psi_a^{(i)}(x_i) \phi_a^{(j)}(x_j)
\eq Kij..
$$
and 
$$
\check{K}_{ij}(x_i, x_j) := K_{ij}(x_i, x_j) - g_{ij}(x_i,x_j),
\eq Kijcheck..
$$
where $g_{ij}(x_i,x_j):= 0$ if $i\le j$.
It may then be shown [EM] that that the probability density \prob may
equivalently be expressed in the form:
$$
P_{N,m} (x^{(j)}_a) =\det (\check{K}_{ij}(x^{(i)}_a, x^{(j)}_b))\ ,
\eq probNm..
$$
where $\check{K}_{ij}(x^{(i)}_a, x^{(j)}_b)$ is viewed as the $((i,a),(j,b))$
element of a matrix of dimension $Nm \times Nm$, labelled by pairs of double
indices
$1 \le i,j \le m, \ 1\le a,b \le N$.
By integrating \probNm over a part of the  variables, it follows [EM] that
the correlation function giving the probability density for finding $k_j$
elements in $\Gamma_j$ at the points $\{x^{(j)}_1, \dots x^{(j)}_{k_j}\}$ for
$j=1 \dots m$ is similary expressed by the 
$\sum_{j=1}^m k_j \times \sum_{j=1}^m k_j$  determinant
$$
\rho_{k_1,\dots k_m} (\{x^{(j)}_1, \dots x^{(j)}_{k_j}\}_{j=1 \dots m}) =\det
(\check{K}_{ij}(x^{(i)}_a, x^{(j)}_b))\vert_{{1\le a \le k_i  \atop 1\le b
\le k_j}}\ .
\eq correlkNm..
$$

 The  functions $K_{ij}$ may also be viewed as kernels of integral operators
$$
\eqalign{
 K_{ij}: \hat{H}_j &\lra \hat{H}_i \cr
  K_{ij}(f)(x_i) &:= \int_{\Gamma_j}K_{ij}(x_i, x_j) f(x_j) d\mu_j(x_j)}
\eq Kij_int..
$$
which, again, map the finite dimensional spaces $\{H_j\}$ to each other. 
Note that the various $K_{ij}$'s may all be obtained from $K_{1m}$ by
composition on the left and right with operators $g_{ij}$:
$$
K_{ij} =  g_{i 1} \circ K_{1m} \circ g_{m j}, \eq Kcomp..
$$
and that, in particular, adjacent ones are related by
$$
K_{ij}= g_{i, i-1}\circ K_{i-1,j} = K_{i, j+1}\circ g_{j+1, j}.  \eq..
$$
It  follows from \dualpsiphi that, when restricted to $H_j$ and $H^j$,
$K_{jj}$ and $K^*_{jj}$ act as identity operators
$$
K_{jj} (\psi) = \psi, \quad \psi\in H_j, \qquad 
K^*_{jj} (\phi) = \phi, \quad \phi  \in H^j.  \eq..
$$

In the following, we denote by $K$,  $g$ and $\check{K}$ the $m \times m$
matrices of integral operators acting on the direct sum 
$$
\hat{H}:= \oplus_{j=1}^m \hat{H}_j, \eq..
$$
with matrix entries $K_{ij}$, $g_{ij}$ and $\check{K}_{ij}$ acting
on the component spaces $\hat{H}_j$.

   Let $\{I_j \ss \Gamma_j\}_{j=1 \dots m}$ be measurable subsets of
the spaces ${\Gamma_j}$, and let $\chi_j$ denote the characteristic
function of $I_j\ss \Gamma_j$. Denote by $\chi_I$ the direct sum of these
functions, viewed as an operator on $\hat{H}$, acting by multiplication by
the various $\chi_j$'s on the component spaces $\hat{H}_j$'s, and 
assume that this leaves $\hat{H}$ invariant. Let 
$$
\check{K}^{\chi_I}:=\check{K}\circ \chi_I \eq..
$$ 
denote the composition of these operators,  also a matrix integral
operator, and let
$$
R^{\chi_I} := ({\bf 1}- \check{K}^{\chi_I})^{-1} \circ \check{K}^{\chi_I}  \eq..
$$
be its Fredholm resolvent. Then the matrix components of this
resolvent, denoted $R_{ij}$, are operators with integral kernels $R_{ij}(x_i,
x_j)$. These also define certain correlation functions, on the product
$\prod_{j=1}^m I_j^N$, the so-called {\it Janossy} densities, given by a 
formula similar to \correlkNm , namely:
$$
\rho^I_{k_1\dots k_m} (\{x^{(j)}_1, \dots x^{(j)}_{k_j}\}_{j=1 \dots m}) =
C_I^{N,m}
\det(R_{ij}(x^{(i)}_a, x^{(j)}_b))\vert_{{1\le a \le k_i  \atop 1\le b
\le k_j}}\ ,
\eq IcorrelkNm..
$$
where the normalization constant $C_I^{N,m}$ is defined to be the Fredholm
determinant:
$$
C_I^{N,m} := \det ({\bf I} -\check{K}^{\chi_I}) \eq freddet..
$$
(which equals the probablity, under the original distribution
\prob, of having no elements within the subset $I=\prod_{j=1}^m I_j^N$).  The
correlation functions \IcorrelkNm  give the probability density of finding,
for $j=1 \dots m$, exactly $k_j$ elements in $I_j$ at the points 
$\{x^{(j)}_1, \dots x^{(j)}_{k_j}\}_{j=1\dots m}$ .

A theorem of Soshnikov [S] expresses this distribution in terms of a new  set
of functions $\{\wt{\psi}^{(j)}, \wt{\phi}^{(j)}, \wt{g}_{ij}\}$, analogous 
to $\{\psi^{(j)},\phi^{(j)}, g_{ij}\}$. To define these functions, one begins
by defining, as in \gijint, integral operators
$$
\wt{g}_{k j} := g_{k,k-1} \circ_{{}_{I_{k-1}^c}} \dots
\circ_{{}_{I_{j+1}^c}}g_{j+1,j}
\eq gkjtildeint..  
$$
where the symbol $\circ_{{}_{I_j^c}}$ denotes composition of integral
operators by integration only over the domain $I_j^c$ complementary to $I_j$
within $\Gamma_j$.  (Note that $\wt{g}_{j+1,j} = g_{j+1.j}$.) Analogously to
\Apairing, we define the pairing matrix
$$
A^I_{ab} := \int_{I^c_1} f_a(x_1) \wt{g}^*_{m1}(h_b)(x_1) d\mu_1(x_1)
=  \int_{I^c_m} \wt{g}_{m1}(f_a)(x_m) (h_b)(x_m) d\mu_m(x_m), 
\eq AIpairing..
$$
and again assume it to be nonsingular.
Once again, making a $PLU$ decomposition of $A^I$ 
$$
A^I := P^I L^I U^I,  \eq..
$$
 we may form  new bases

$$
\wt{\psi}_a^{(1)} := \sum_{b=1}^N(P^IL^I)^{-1} _{ab} f_b \qquad
\wt{\phi}_a^{(m)} :=
\sum_{a=1}^N(U^I)^{-1} _{ba} h_a 
\eq..
$$
for the spaces $H_1$ and $H^m$, as well as bases for the
sequence of dual pairs of spaces  obtained by composition with
$\wt{g}_{j 1}$ and $\wt{g}^*_{mj}$:
$$
\wt{\psi}_a^{(j)}:=\wt{g}_{j1} (\wt{\psi}_a^{(1)}), \qquad
\wt{\phi}_a^{(j)}:=\wt{g}^*_{mj} (\wt{\phi}_a^{(m)}).
\eq defwtpsiphi..
$$
These are mutually dual in the sense that
$$
\int_{I^c_j} \wt{\psi}_a^{(j)}(x_j) \wt{\phi}_b^{(j)}(x_j)
d\mu_j(x_j) =
\delta_{ab}.
\eq dualpsiphitilde..
$$

Define, as before, a set of integral kernels:
$$
\wt{K}_{ij}(x_i, x_j) := \sum_{a=1}^N \wt{\psi}_a^{(i)}(x_i)
\wt{\phi}_a^{(j)}(x_j)
\eq Ktildeij..
$$
and 
$$
\check{\wt{K}}_{ij}(x_i, x_j) :=\wt{K}_{ij}(x_i, x_j) - \wt{g}_{ij}(x_i,x_j),
\eq Ktildecheckij..
$$
where again, $\wt{g}_{ij}(x_i,x_j):= 0$ if $i\le j$, and denote by
$\wt{K}_{ij}$,  $\wt{g}_{ij}$ and $\check{\wt{K}}_{ij}$ the corresponding
integral operators mapping $\hat{H_j} \ra \hat{H}_i$.  The associated
$m \times m$ matrix integral operators acting on $\hat{H}$ are similarly
denoted  $\wt{K}$, $\wt{g}$ and $\check{\wt{K}}$, with
$$
\check{\wt{K}} := \wt{K} - \wt {g}.  \eq Kminusg..
$$
Then Soshnikov's theorem states that the resolvent operator $R^{\chi_I}$
entering in the definition of the Janossy distribution coincides with the
operator $\check{\wt{K}}{}^{\chi_I} := \check{\wt{K}}{} \circ \chi_I$  so
constructed.
\Proclaim Theorem 1:
$$
  R^{\chi_I} = \check{\wt{K}}{}^{\chi_I}.  \eq..
$$

  A proof of this theorem is given in [S], based on decomposition of the spaces
into various subspaces, and an inductive construction, but it is rather long
and intricate. In the following section, a very simple direct proof will
be given.

Since no further effort is required, we will actually prove an obvious
generalization that reduces to the above result as a special case.
Namely, instead of choosing measurable subsets $I_j\ss \Gamma_j$ and their
characteristic functions $\chi_j$, we may replace the latter by any set of
square integrable functions $\{w_j \in L^2(\Gamma_j, d\mu_j)\}_{j=1 \dots m}$,
and generalize the definition of the matrix $A^I$ and the functions
$\{\wt{\psi}^{(j)}, \wt{\phi}^{(j)},\wt{g}_{ij}\}$ accordingly to:
$$
\eqalign{
A^w_{ab} &:= \int_{\Gamma_1} f_a(x_1) \wt{g}^*_{m1}(h_b)(x_1)
(1-w_1(x_1))d\mu_1(x_1) \cr
&=  \int_{\Gamma_m} \wt{g}_{m1}(f_a)(x_m) (h_b)(x_m)
(1-w_m(x_m))d\mu_m(x_m),} 
\eq AIpairingnew..
$$
with $PLU$ decomposition
$$
A^w := P^w L^w U^w, \eq..
$$
where, denoting by ${}\circ_{v_j}$ the operation of mutiplication by a 
function $v_j$ in $L^2(\Gamma_j, d\mu_j)$ followed by composition, we 
replace the definition of $\wt{g}_{k, j}$ in \gkjtildeint by
$$
\wt{g}_{k j} := g_{k,k-1}\circ_{1-w_{k-1}} \circ \dots
 \circ_{1- w_{j+1}}\circ  g_{j+1,j}.
\eq gkjtildeintnew..  
$$
Then
$$
\wt{\psi}_a^{(1)} := \sum_{a=1}^N(P^w L^w)^{-1} _{ab} f_a \qquad
\wt{\phi}_a^{(m)} :=
\sum_{b=1}^N(U^w)^{-1} _{ba} h_a, 
\eq..
$$
define bases for the spaces $H_1$ and $H^m$, respectively, while bases for
the sequence of dual spaces obtained by composition with $\wt{g}_{j, 1}$ and
$\wt{g}^*_{m,k}$, as defined in \gkjtildeintnew, are again given by
\defwtpsiphi, with the dualization pairing given by
$$
\int_{\Gamma_j} \wt{\psi}_a^{(j)}(x_j) \wt{\phi}_b^{(j)}(x_j)(1-w_j(x_j))
d\mu_j(x_j) =
\delta_{ab}.
\eq dualpsiphitildew..
$$
  Replacing the characteristic functions $\chi_I = (\chi_1 \dots \chi_m)$  by
the set of weight functions  $w:= (w_1,\dots w_m)$, the definitions of the
integral operators and kernels $\wt{K},\wt{g}, \check{\wt{K}}$  retain the 
same form and the theorem becomes
\Proclaim Theorem 2:
$$
  R^w = \check{\wt{K}}{}^w := \check{\wt{K}}\circ w .  \eq RwKweq..
$$
where
$$
R^w := (1- \check{K}{}^w)^{-1} \circ \check{K}{}^w  \eq Rwdef..
$$
is the Fredholm resolvent of
$$
\check{K}{}^w{} := \check{K}\circ w. \eq..
$$

Such a generalization is of use  in multimatrix models, since it
allows us to replace, for example, the characteristic functions $\chi_j$ on
subintervals $I_j\ss \Gamma_j$ by weighted characteristic functions
$\sum_{\alpha=1}^{k_j}\kappa_{\alpha, j}\chi_{I_{\alpha, j}}$  over unions of
disjoint subintervals $I_j = \cup_{\alpha=1}^{k_j} I_{\alpha, j}$. The
corresponding  Fredholm determinant \freddet, expanded as a power series in 
the coefficients $\kappa_{\alpha, j}$ becomes a generating function for the
probabilities of having given numbers of eigenvalues within the subintervals,
as in the
$1$-matrix case [TW].

%%%%%%%%%%%%%%%%%%%%%%%%%%%% Section 2. Proof %%%%%%%%%%%%%%%%%%%
\section 2. Proof of the theorem.
\smallskip \nobreak

The equality \RwKweq may equivalently be written as
$$
\check{K}^w \circ \check{\wt{K}}{}^w = \check{\wt{K}}{}^w - \check{K}^w, 
\eq KtimesKcheck..
$$
or, explicitly,
$$
\sum_{j=1}^m \check{K}_{ij} \circ_{w_j} \check{\wt{K}}_{jk} = 
\check{\wt{K}}_{ik} - \check{K}_{ik} \eq compKtildeK..
$$
(where composition on the right  by $w$ is omitted, since the equality will
be shown to hold without it). Substituting  \Kijcheck and \Kminusg  in \KtimesKcheck, this 
is equivalent to
$$
K \circ_w \wt{K} -  g \circ_w \wt{K} - K \circ_w \wt{g}
+ g \circ_w \wt{g}  + \wt{g} - g = \wt{K} - K{}.  \eq compKtildeKexp..
$$
This relation follows as a consequence of four identities relating the
various summands on the left, each of which is easily proved:

$$
g \circ_w \wt{g}  + \wt{g} - g = 0 \eq gtildeg..
$$
$$
(K \circ_w \wt{K})_{ij} = g_{i1}\circ\wt{K}_{1j} +\d_{i1} \wt{K}_{1j} -
K_{im}\circ_{1-w_m}\wt{g}_{mj} - K_{im}\d_{mj}
\eq KtildeK..
$$
$$
(g \circ_w \wt{K})_{ij} = g_{i1}\circ \wt{K}_{1j} 
+\d_{i1} \wt{K}_{1j} - \wt{K}_{ij}
\eq gtildeK..
$$
$$
(K \circ_w \wt{g})_{ij} =  K_{ij} - K_{im}\circ_{1-w_m}\wt{g}_{mj}
- K_{im}\d_{mj}
\eq Ktildeg..
$$

\Demo{Proof of \gtildeg}: From \gijint and \gkjtildeintnew we have, for  
$k -1 > j >i+1$, 
$$
\eqalign{
g_{kj} \circ_{w_j} \wt{g}_{ji} 
&= g_{kj} \circ \wt{g}_{ji} - g_{kj} \circ_{1-w_j} \wt{g}_{ji} \cr
&= g_{kj} \circ \wt{g}_{ji} - g_{k,j+1}\circ g_{j+1,j}
\circ_{1-w_j}\wt{g}_{ji}
\cr &= g_{kj}\circ \wt{g}_{ji} - g_{k, j+1} \circ
\wt{g}_{j+1,i},} \eq..
$$
while for $j=k-1$,
$$
\eqalign{
g_{k, k-1} \circ_{w_{k-1}} \wt{g}_{k-1,i}
& = g_{k, k-1} \circ \wt{g}_{k-1,i} -
g_{k, k-1} \circ_{1-w_{k-1}} \wt{g}_{k-1,i} \cr
&= g_{k, k-1}\circ \wt{g}_{k-1,i} -
\wt{g}_{ki},}\eq..
$$
and for $j=i+1$
$$
\eqalign{
g_{k, i+1} \circ_{w_{i+1}} \wt{g}_{i+1,i} 
& = g_{k, i+1}\circ \wt{g}_{i+1,i} 
-g_{k, i+1} \circ_{1-w_{i+1}} \wt{g}_{i+1,i} \cr
&= g_{ki} -g_{k, i+2} \circ \wt{g}_{i+2,i} .} \eq..
$$

Summing over $j$ and cancelling all intermediate terms  gives the
result
$$
(g\circ_w\wt{g})_{ki}=\sum_{j=i+1}^{k-1} g_{kj}\circ_{w_j}
\wt{g}_{ji} =g_{ki} - \wt{g}_{ki}.
\eq..
$$
(Note that this means that the integral operator $g^w$ is the Fredholm
resolvent of $\wt{g}{}^w$.)
\Demo{Proof of \KtildeK}: 
For $j\neq m$, we have
$$
\eqalign{K_{ij} \circ_{w_j}\wt{K}_{jk} 
&= K_{ij} \circ \wt{K}_{jk} - K_{ij} \circ_{1-w_j} \wt{K}_{jk} \cr 
&= K_{ij} \circ \wt{K}_{jk} - K_{i,j+1}\circ g_{j+1,j} \circ_{1-w_j} 
\wt{K}_{jk} \cr
&=  K_{ij} \circ \wt{K}_{jk} -  K_{i,j+1} \circ \wt{K}_{j+1,k}.}
\eq..
$$
Therefore, summing and cancelling the intermediate terms gives
$$
\sum_{j=1}^m K_{ij} \circ_{w_j}\wt{K}_{jk} =
K_{i1} \circ\wt{K}_{1k} - K_{im} \circ_{1-w_j}\wt{K}_{mk}.
\eq KtilkdeKprep..
$$

By \dualpsiphi and \dualpsiphitildew, the operators $K_{11}$ and
$(\wt{K}_{mm}{}\circ (1-w_m){})^*$ act as the identity on the spaces $H_1$ 
and
$H^m$, respectively, 
and hence
$$
K_{11}\circ \wt{K}_{1k} = \wt{K}_{1k}, \qquad K_{im}\circ_{1-w_m} \wt{K}_{mm} =
\wt{K}_{im} .  \eq..
$$
It follows from  \Kcomp and the corresponding relations
$$
\wt{K}_{ij} =  \wt{g}_{i 1} \circ_{1-w_1} \wt{K}_{1m} \circ_{1-w_m}
\wt{g}_{m j} 
\eq Ktildecomp..
$$
for the $\wt{K}_{ij}$'s that, for $i \neq 1$, $j\neq m$,
$$
K_{i1}\circ \wt{K}_{1k} = g_{i1}\circ\wt{K}_{ik}, \qquad K_{im}\circ_{1-w_m}
\wt{K}_{mj} =
K_{im}\circ_{1-w_m} \wt{g}_{mj}.  \eq..
$$
Combining these relations with \KtilkdeKprep leads to \KtildeK.

\Demo{Proof of \gtildeK}:  For $j<i-1$, we have
$$
\eqalign{
g_{ij} \circ_{w_j} \wt{K}_{jk} &= g_{ij}\circ \wt{K}_{jk} 
- g_{ij}\circ_{1-w_j}\wt{K}_{jk} \cr
&=g_{ij}\circ \wt{K}_{jk} 
- g_{i,j+1}\circ g_{j+1,j} \circ_{1-w_j}\wt{K}_{jk} \cr
 &= g_{ij}\circ\wt{K}_{jk} - g_{i,j+1} \circ \wt{K}_{j+1,k},}
\eq..
$$
while for $j=i-1$,
$$
\eqalign{
g_{i, i-1} \circ_{w_{i-1}} \wt{K}_{i-1,k} &= g_{i,i-1}\circ \wt{K}_{i-1,k} -
g_{i, i-1}
\circ_{1-w_{i-1}}
\wt{K}_{i-1,k} \cr
 &= g_{i,i-1} \circ \wt{K}_{i-1,k} -  \wt{K}_{ik}, }
\eq..
$$
Again, summing, and cancelling the intermediate terms gives:
$$
\eqalign{\sum_{j=1}^mg_{ij} \circ_{w_j} \wt{K}_{jk} =
\sum_{j=1}^{i-1}g_{ij}  \circ_{w_j} \wt{K}_{jk} 
&= g_{i1}\circ\wt{K}_{1k}-\wt{K}_{ik},}\eq..
$$
which is \gtildeK (the case $i=1$  being trivially satisfied).

\Demo{Proof of \Ktildeg}:  For $m>j>k+1$, we have
$$
\eqalign{
K_{ij} \circ_{w_j} \wt{g}_{jk} &= K_{ij}\circ \wt{g}_{jk} - 
K_{ij} \circ_{1-w_j}\wt{g}_{jk} \cr
&= K_{ij}\circ \wt{g}_{jk} - 
K_{i,j+1}\circ g_{j+1,j} \circ_{1-w_j}\wt{g}_{jk} \cr
 &= K_{ij}\circ\wt{g}_{jk} - K_{i,j+1} \circ \wt{g}_{j+1,k},}
\eq..
$$
while for $j=k+1$,
$$
\eqalign{
K_{i, k+1} \circ_{w_{k+1}} \wt{g}_{k+1,k} &= K_{i,k+1}\circ \wt{g}_{k+1,k} -
K_{i, k+1}\circ_{1-w_{k+1}} \wt{g}_{k+1,k} \cr
 &=  K_{ik}  - K_{i,k+2} \circ \wt{g}_{k+2,k},}
\eq..
$$
Again, summing and cancelling the intermediate terms gives:
$$
\eqalign{\sum_{j=1}^m K_{ij} \circ_{w_j} \wt{g}_{jk} =
\sum_{j=k+1}^m K_{ij}  \circ_{w_j} \wt{g}_{jk} 
&= K_{ik} - K_{im}\circ_{1-w_m}\wt{g}_{mk},}\eq..
$$
which is \Ktildeg (the case $j=m$   again being trivially satisfied).

Combining the four relations \gtildeg, \KtildeK, \gtildeK and \Ktildeg
gives \compKtildeKexp, hence proving the theorem.

 \noindent{\it Remark.} Although this theorem has been formulated with respect to probability measures and Hilbert spaces of functions on them, it may clearly be extended to more general cases involving, e.g., complex measures, provided all the integrals appearing are convergent, and the related integral operators and their composites are well-defined.

\bigskip \bigskip

%%%%%%%%%%%%%%%%%% Acknowledgements  %%%%%%%%%%%%%%%%%%%
\bigskip\bigskip \noindent{\it Acknowledgements.}
The author would like to thank M. Bertola and A. Soshnikov for helpful discussions. This research 
 was supported in part by the Natural Sciences and Engineering Research Council of Canada.
\bigskip \bigskip
%%%%%%%%%%%%%%%%%% References %%%%%%%%%%%%%%%%%%%%%%%%
\goodbreak
\references

BEH1& M. Bertola, B. Eynard, J. Harnad, ``Duality, 
Biorthogonal Polynomials and Multi--Matrix Models'', 
{\it  Commun. Math. Phys.} {\bf 229},  73-120 (2002).

BEH2& M. Bertola, B. Eynard, J. Harnad,``Differential systems for 
biorthogonal polynomials appearing in 2-matrix models, and the associated 
Riemann-Hilbert problem'',  {\it  Commun. Math. Phys.}, , {\bf 243}, 
193-240 (2003).
 
BS& A. Borodin and A. Soshnikov, ``Janossy densities I. Determinantal
ensembles'', {\it J. Stat. Phys} {\bf 113} 611--622 (2003).

EM& B. Eynard  and M.~	L.  Mehta, ``Matrices coupled in a chain I. 
Eigenvalue correlations'', {\it J. Phys. A: Math. Gen.} {\bf 31}, 4449
(1998).

J& K. Johannson, ``Discrete polynuclear growth and determinantal processes'',
\hfil \break arXiv:math.PR/0206028.

S& Alexander Soshnikov, ``Janossy densities of coupled random matrices'',
{\it  Commun. Math. Phys.}  (2004, to appear), arXiv:math-ph/0309019.

PS& M. Pr\"ahofer and H. Spohn, ``Scale invariance of the PNG droplet and 
the Airy process'', {\it J. Stat. Phys.} {\bf 108} 1071--1106 (2002).

TW&  Craig A. Tracy, and Harold Widom,  ``Correlation functions,
cluster functions,  and spacing distributions for random
matrices'', {\it J. Stat. Phys.} {\bf  92},  809--835 (1998).

\endreferences

\end